# Enhanced Thermoelectric Properties of Dirac Semimetal $Cd_3As_2$

*Tong Zhou[1,2], Cheng Zhang[1,3], Huisheng Zhang[1,2], Faxian Xiu[1,3], and Zhongqin Yang[1,2,3]\**

1 State Key Laboratory of Surface Physics and Department of Physics, Fudan University, Shanghai 200433, China

2 Key Laboratory for Computational Physical Sciences (MOE), Fudan University, Shanghai 200433, China

3 Collaborative Innovation Center of Advanced Microstructures, Fudan University, Shanghai, 200433, China

\*Address correspondence to: zyang@fudan.edu.cn

## ABSTRACT

We report an investigation of temperature- and doping-dependent thermoelectric behaviors of topological semimetal $Cd_3As_2$. The electrical conductivity, thermal conductivity, Seebeck coefficient, and figure of merit (*ZT*) are calculated by using Boltzmann transport theory. The calculated thermoelectric properties of the pristine $Cd_3As_2$ match well the experimental results. The electron or hole doping, especially the latter, is found improving much the thermoelectric behaviors of the material. The optimum merit *ZT* of $Cd_3As_2$ with electron doping is found to be about 0.5 at T=700 K with n=1×10$^{20}$ cm$^{-3}$, much larger than the maximum experimental value obtained for the pristine $Cd_3As_2$ (~0.15). For the p-type $Cd_3As_2$, the maximal value of the Seebeck coefficient as a function of temperature increases apparently with the increase of the hole doping concentration and its position shifts drastically towards the lower temperature region compared to that of the n-type $Cd_3As_2$, leading to the optimum merit *ZT* of about 0.5 obtained at low temperature of 500K (p=1×10$^{20}$ cm$^{-3}$) in the p-type $Cd_3As_2$.

# I. INTRODUCTION

Thermoelectric performance is of considerable importance in industry due to its numerous potential applications, including vehicular exhaust waste-heat recovery, energy harvesting, heating and cooling, and solid-state energy conversion[1,2]. High thermoelectric performance is extremely desired for these applications[3,4]. Thermoelectric behaviors are typically quantified in term of a dimensionless figure of merit (*ZT*) given by the following expression: $ZT = \frac{S^2 \sigma T}{\kappa}$, in which *S* is the Seebeck coefficient or thermopower, σ is the electrical conductivity, T is the absolute temperature, and κ is the thermal conductivity, normally the sum of electronic and lattice contributions, namely $\kappa = \kappa_e + \kappa_L$. The $S^2\sigma$ is called power factor of the thermoelectric performance. The *ZT* expression shows that for good thermoelectric performance both the Seebeck coefficient and electrical conductivity are expected to be high, while the thermal conductivity is expected to be low. These parameters are, however, intimately coupled to each other, providing conflicts in the optimization process[1,2]. Hence, optimizing and finding a usable high-performance thermoelectric material remain full of challenges.

Recently, three-dimensional (3D) Dirac semimetal states have been theoretically predicted and experimentally realized in $Cd_3As_2$[5-9]. Unlike other semimetals, the $Cd_3As_2$ crystal possesses Dirac fermions that disperse linearly in k-space and as a result become one of the crystalline materials with a ultrahigh electron mobility μ of about $10^4$-$10^6$ $cm^2V^{-1}s^{-1}$[8,9]. Since the power factor strongly depends on the electron mobility i.e., $S^2\sigma \approx \mu (m^*/m_e)^{1.5}$ (where m* is the energy-band electron effective mass and $m_e$ is the free electron mass)[10], $Cd_3As_2$ shows a great potential for the high performance of thermoelectric applications. The maximum *ZT* value of the pristine $Cd_3As_2$ achieved in experiments is, however, only about 0.15[11,12]. How to improve the thermoelectric performance in this material is of significance. For semimetals, they usually can be easily doped with both electrons and holes due to the existence of the Dirac cones near the Fermi level ($E_F$), which has been proved in graphene[13]. Both the n-type[11,12,14,15] and p-type[16,17] $Cd_3As_2$ have also been successfully synthesized in experiments. It is meaningful to explore systematically the effect of the electron or hole doping on the thermoelectric properties of the semimetal $Cd_3As_2$.

In this work, we perform first principles computations on the electronic states and thermoelectric properties of the topological semimetal $Cd_3As_2$. It is found that the electronic structure of $Cd_3As_2$ can be described correctly by generalized gradient approximation after the spin-orbit coupling is considered. Based on Boltzmann transport theory, we calculate the electrical conductivity, thermal conductivity, Seebeck coefficient, and figure of merit of the semimetal. The calculated thermoelectric properties of the pristine $Cd_3As_2$ match well the experimental results. Very interestingly, the *ZT* value of the semimetal is found being improved much by both electron or hole doping, especially the latter. For the n-type $Cd_3As_2$, the optimal doping level is about $1\times10^{20}$ cm$^{-3}$, leading to the *ZT* increasing to about 0.5 at T=700 K, much larger than the *ZT* (0.15) of the pristine $Cd_3As_2$. Differently, for the p-type $Cd_3As_2$, the maximal value of Seebeck coefficient shifts towards the low temperature region and the *S* value in the hole-doping region is much larger than the one in the electron-doping region, resulting in the optimum *ZT* of about 0.5 at 500 K for the p-type $Cd_3As_2$. Our work provides new paths towards high thermoelectric performance in the topological Dirac semimetals.

## II. METHODS AND MODELS

The calculations are performed by using the general potential linearized augmented plane-wave (LAPW) method[18] as implemented in the WIEN2K code[19]. Spin-orbital coupling is included as a second vibrational step by using scalar-relativistic eigenfunctions as the basis after the initial calculation is converged to self-consistency. The convergence of the calculations regarding the size of the basis set is achieved by using an $R_{MT}K_{max}$ value of 7 where $R_{MT}$ is the smallest atomic sphere radius in the unit cell and $K_{max}$ is the magnitude of the largest K wave vector inside the first Brillouin zone. Exchange and correlation effects are accounted for by using the generalized gradient approximation Perdew-Burke-Ernzerh-of (PBE) functional[20]. Considering the possible underestimation of the band gap by PBE, we do the calculation based on modified Becke-Johnson (mBJ) potential[21], which has been proved to be a good correction to the exchange and correlation functional for electronic structure calculations of thermoelectric systems[22-24]. From the calculated band structure, the thermoelectric transport tensors are evaluated using the semi-classical Boltzmann kinetic transport theory within the constant relaxation time approximation (CSTA) and the rigid band approach as implemented within the BoltzTraP program[25]. The constant relaxation time is given as a standard electron-phonon model[22,26], the parameter of which is given by fitting the experimental data. This method has been successfully described the transport coefficients of a wide range of thermoelectric materials[27–32].

## III. RESULTS AND DISCUSSION

$Cd_3As_2$ has a distorted superstructure of the antifluorite ($M_2X$) structure type with an $I4_{1/acd}$ space group[5]. It can be viewed as a tetragonally-distorted anti-fluorite structure with 1/4 M sites of vacancies and containing 80 atoms per unit cell. The calculated electronic band structures without and with spin-orbit coupling (SOC) based on the PBE potentials are shown in Fig. 1 (a) and (b), respectively. As shown in Fig. 1(a), the conduction and valence bands of $Cd_3As_2$ are degenerate at the $\Gamma$ point without the SOC. Similar to most of the semiconductors with antifluorite orzinc-blende structures, the low energy electronic properties of $Cd_3As_2$ are mostly determined by the Cd 5s states (conduction bands) and the As 4p states (valence bands). When the SOC is considered, the bands are inverted around the $\Gamma$ point with the s states lower than the p states, which is an important sign of the nontrivial topology appearance. Different from the ordinary topological insulators, no global band gap opens in the $Cd_3As_2$ even after the SOC is taken into account[5,8]. Instead the bands cross around the $\Gamma$ point along the $\Gamma$-Z direction, exactly at the $E_F$, due to the protection of the $C_{4v}$ symmetry with respect to the $k_z$ axis. This band unique characteristic together with the time-reversal and space inversion symmetries gives rise to a 3D massless Dirac semimetal of the system, in good agreement with the results of Refs.[5,8]. Since material thermoelectric properties strongly depend on the band structure[25] and the mBJ potential was proved to be an excellent method for the electric states of the thermoelectric systems[22-24], we do the calculation for $Cd_3As_2$ based on mBJ potential. The electronic band structures without and with SOC based on the mBJ potential are given in Fig. 1(c) and (d), respectively. A global gap of 160 meV is opened at the $\Gamma$ point in Fig. 1(d), inconsistent with the experimental data[8,9], indicating the mBJ potential cannot describe well the electronic structure of 3D Dirac semimetal and even produce a wrong result. This observation might be explained by the fact that the mBJ potential was designed to reproduce the shape of the exact exchange optimized effective potential of atoms[21,22,24]. Because mBJ potential leads to an increase of the gap values, the conduction bands and valances band are separated too much, making the Dirac point disappear. It can be expected that if the gap opened by the mBJ in Fig. 1(c) is not so much and the band inversion can still be induced by the SOC interaction, the 3D Dirac semimetal behavior will keep in the system. The reason is that the mBJ potential does not break the $C_{4v}$ symmetry in the system.

Since the electronic structure of $Cd_3As_2$ is described correctly with the PBE functional, the thermoelectric properties of $Cd_3As_2$ are investigated by using Boltzmann transport theory based on the band structures obtained with the PBE functional (Fig. 1(b)). In experiments, the fabricated intrinsic $Cd_3As_2$ is an n-type material with an electron carrier concentration of about $n=1\times10^{19}$ cm$^{-3}$[11,12], also called the pristine $Cd_3As_2$ in this work. Thus, we first calculate the thermoelectric properties of $Cd_3As_2$ at this electron carrier concentration. Within the framework of the Boltzmann transport theory, the scattering time relaxation ($\tau$) is usually adopted approximately as a constant, which is associated with the behaviors of the electrical conductivity ($\sigma$), thermal conductivity from electronic contributions ($k_e$), and further the *ZT* of the system[25]. The relaxation time generally depends on both the charge carrier concentration (n) and the temperature (T). We here employ a standard electron-phonon dependence on *T* and n for $\tau$, namely, $\tau = \dfrac{C}{Tn^3}$[22,26], where C is a constant and can be determined by comparing to experimental data. For this pristine $Cd_3As_2$ sample (with $n=1\times10^{19}$ cm$^{-3}$), the experimental electrical conductivity is about 67 S/cm at 300 K[11,12], which together with the ratio of $\sigma/\tau$ obtained from the Boltzmann transport theory gives the C to be about $5\times10^{-6}$ sKcm.

The calculated electrical resistivity of the pristine $Cd_3As_2$ sample with respect to the temperature is plotted in Fig. 2(a). It shows a metallic behavior at low temperatures, in good agreement with the experiments, indicating that the standard electron-phonon model is suitable to describe the mechanism of electron-phonon scattering in $Cd_3As_2$. Note that since the experimental data of $Cd_3As_2$ range from 0

K to 380 K, we plot the thermoelectric properties obtained in our calculations also in this temperature range in Fig. 2 to compare with the experimental results directly. With the constant relaxation time approximation, the Seebeck coefficient is independent of τ. The obtained Seebeck coefficient of the $Cd_3As_2$ sample also matches the experiment very well (Fig. 2(b)), meaning the semiclassical Boltzmann transport theory can be adopted to describe the thermoelectric transport properties of $Cd_3As_2$. The calculated electron thermal conductivity ($\kappa_e$), plotted in Fig. 2(c), increases with the temperature. To evaluate the ZT of this $Cd_3As_2$ sample, the lattice thermal conductivity ($\kappa_L$) is also required, which is obtained by subtracting $\kappa_e$ from the total thermal conductivity ($\kappa_{tot}$) provided in experiments[11]. We find the obtained $\kappa_L$ follows a classic A/T dependence with A= 245 W/m, as shown in Fig. 2(c). At high temperatures, $\kappa_L$ decreases, giving rise to $\kappa_e$ contributing more to the total thermal conductivity. Based on the obtained σ, S, and $\kappa_{tot}$, the thermoelectric figure of merit can be estimated through the formula $ZT = S^2\sigma T/\kappa_{tot}$ (Fig. 2(d)). The trend of the calculated ZT, especially at low temperature region, is in good agreement with the experimental results[11,12]. The maximum of the calculated ZT is about 0.2 at 400 K, also very close to the experimental values[11,12]. These good agreements indicate that Boltzmann transport theory with constant relaxation time approximation can describe very well the thermoelectric transport properties of Dirac semimetals.

To investigate the thermoelectric properties of $Cd_3As_2$ with different electron carrier concentrations, the rigid-band approach is employed. Fig. 3 shows the evolution of the electrical conductivity with respect to the temperature for various electron carrier concentrations of interest. At a fixed temperature, the electrical conductivity increases drastically with the carrier concentration. When concentrations is n= $1\times10^{21}$ cm$^{-3}$, the electrical conductivity is up to 3.81/uΩm at 300 K, which is much higher than that of the pristine $Cd_3As_2$ (2.62/uΩm at 300 K). The calculated Seebeck coefficient S (Fig. 3(b)) has, however, different trends as a function of the temperature for various electron carrier concentrations. When the electron concentration is low (n ≤ $5\times10^{19}$ cm$^{-3}$ ), the absolute value of the calculated S (|S|) increases with the temperature and then decreases. The maximal value of |S| increases with the concentration and its temperature position shifts up. For high electron concentrations (n >$1\times10^{20}$ cm$^{-3}$),

the calculated |S| increases almost linearly with the temperature in the range below 900 K, like a metal[1]. As a consequence, for the n-type doping, the maximum of the calculated |S| is about 170 µV/K with n = $5\times10^{19}$ cm$^{-3}$ at 700 K. Combining the electron thermal conductivity and lattice thermal conductivity, we plotted the total thermal conductivity with respect to the temperature at various concentrations in Fig. 3(c). Since the $\kappa_L$ decreases drastically with the temperature ($\kappa_L = A/T$), the trend of $\kappa_{tot}$ is determined by the $\kappa_e$ for the temperatures larger than 200 K. With the obtained σ, S, and $\kappa_{tot}$, the figure of merit of the material at various electron concentrations can be calculated, presented in Fig. 3(d). Obviously, the ZT shares the similar tendency with |S| as a function of concentration (Fig. 3(b)) to certain extent. The maximum optimum ZT of Cd$_3$As$_2$ is found to be about 0.5 at T = 700 K with n = $1\times10^{20}$ cm$^{-3}$. This carrier concentration is hopefully achieved in experiments with current advanced technologies[11,16,33]. This predicted ZT value (0.5) of Cd$_3$As$_2$ with electron doping is much larger than the maximum ZT (0.15) achieved in the experiments for the pristine samples[11,12].

The thermoelectric behavior of Cd$_3$As$_2$ with hole doping is also explored since electron and hole doping are both easily carried out in semimetals in experiments[16,17]. As shown in Fig. 4(a) and (c), both the trends of electronic conductivity and thermal conductivity of the p-type Cd$_3$As$_2$ as functions of the temperature and concentration are similar to those of the n-type Cd$_3$As$_2$, respectively. The magnitudes of σ and $\kappa_{tot}$ in the p-type Cd$_3$As$_2$ are, however, usually lower than those of the n-type Cd$_3$As$_2$ at a fixed temperature or concentration, respectively, associated with the band structures. The S curves in the p-type Cd$_3$As$_2$ (Fig. 4(b)) is much different from that of the n-type Cd$_3$As$_2$ (Fig. 3(b)). Fig. 4(b) shows that the maximal values of S at various hole concentrations are located at the lower temperatures than those of the n-type Cd$_3$As$_2$ and the S absolute values in the hole-doping region are much larger than the corresponding ones in the electron-doping region. As a result, the ZT of the p-type Cd$_3$As$_2$ is enhanced much for the middle temperature region (Fig. 4(d)), compared to that of the n-type sample. For the p-type Cd$_3$As$_2$, an optimum ZT of about 0.5 can be acquired at 500 K with a hole doping concentration of p=$1\times10^{20}$ cm$^{-3}$, as shown in Fig. 4(d).

Because the concentrations of Cd$_3$As$_2$ are usually tuned by the gate voltage[16,17], it is significant to give the thermoelectric properties with respect to chemical potential at various temperatures. Thus, we plot the σ, S, $\kappa_{tot}$, and *ZT* in terms of chemical potential at various temperatures as Fig. S1(a)-(d) in the Electronic supplementary information, respectively. At the same temperature, the maximum value of the σ with electron doping is larger than that with hole doping and the σ of both n-type and p-type $Cd_3As_2$ decreases with the temperature (Fig. S1(a)), consistent with the trends in Figs. 3(a) and 4(a). For the Seebeck coefficient, it is interesting to find that its maximum values of the hole doping at different temperatures all occur at about 0.1 eV below the $E_F$, where the band dispersion is weak (Fig. 1(b)). While for the electron doping, the S maximum values all occur at about 0.25 eV above the $E_F$ due to the degenerated bands around this energy (Fig. 1(b)). For most temperatures, the S maximum values of the p-type $Cd_3As_2$ are larger than those of the n-type $Cd_3As_2$, resulting in a larger *ZT* in p-type $Cd_3As_2$, also in agreement with the trends obtained from Fig. 3(d) and 4(d). The reason why the thermoelectric behavior of the p-type $Cd_3As_2$ is superior to that of the n-type $Cd_3As_2$ can be ascribed to the band dispersions around the $E_F$. The relation between temperature and Seebeck coefficients can be seen from relatively simple models of electron transport. For metals or degenerate semiconductors, the Seebeck coefficient is approximately given by[1]:

$$S = \frac{8\pi^2 k_B^2 T}{3qh^2} m^* (\frac{\pi}{3n})^{2/3},$$

where *q* is the carrier charge, *n* is the carrier concentration, and *m\** is the effective mass of the carrier. As shown in Fig. 1(b), the bands around the valence band maximum (VBM) disperse less than those around the conduction band minimum (CBM) do. A weak dispersion band gives the electrons with heavy effective masses, which can enhance the Seebeck coefficient[1]. This trend can also be obtained by the analysis of the densities of states (DOSs) of $Cd_3As_2$. As shown in the Fig. S2 in the Electronic supplementary information, the slope of the total DOS at the VBM is larger than that at the CBM, indicating that the effective mass at the VBM is larger than that at the CBM, in agreement with the band analysis. Since the m* of p-type is larger than that of n-type near the $E_F$, the maximal values of S at various hole concentrations are located at the lower temperatures than those of the n-type $Cd_3As_2$. This trend can also be found in other thermoelectric materials, such as PbTe[34] and $CuCoO_2$[35].

## IV. CONCLUSIONS

In summary, we performed the first principles computations on the electronic structure and thermoelectric behaviors of the topological semimetal $Cd_3As_2$ and found that the electronic structure of $Cd_3As_2$ can be described correctly by generalized gradient approximation with spin-orbit coupling. Based on the Boltzmann transport theory, the electrical conductivity, thermal conductivity, Seebeck coefficient, and figure of merit of the system were calculated. The optimum *ZT* of $Cd_3As_2$ with electron-doping was found to be about 0.5 at T=700 K with n=$1\times10^{20}$ cm$^{-3}$, while for the p-type $Cd_3As_2$, the maximal value of Seebeck coefficient as a function of temperature increases with the increase of the hole doping concentration and its position shifts drastically towards the lower temperature region compared to that of the n-type $Cd_3As_2$, leading to high *ZT* (0.5) of the p-type $Cd_3As_2$ achieved at low temperature (500 K). Our work provides a new avenue towards high-performance thermoelectric materials based on topological Dirac semimetals.

**ACKNOWLEDGMENTS**

This work was supported by National Natural Science Foundation of China with Grant No. 11574051, Natural Science Foundation of Shanghai with Grant No. 14ZR1403400, and Fudan High-end Computing Center.

**Electronic supplementary information (ESI) available:** Calculated electrical conductivity (σ), Seebeck coefficient (*S*), thermal conductivity (κ), and figure of merit (*ZT*) of n/p-type $Cd_3As_2$ with respect to the chemical potential at different temperatures and calculated densities of states for $Cd_3As_2$ crystal.

**Figures and captions**

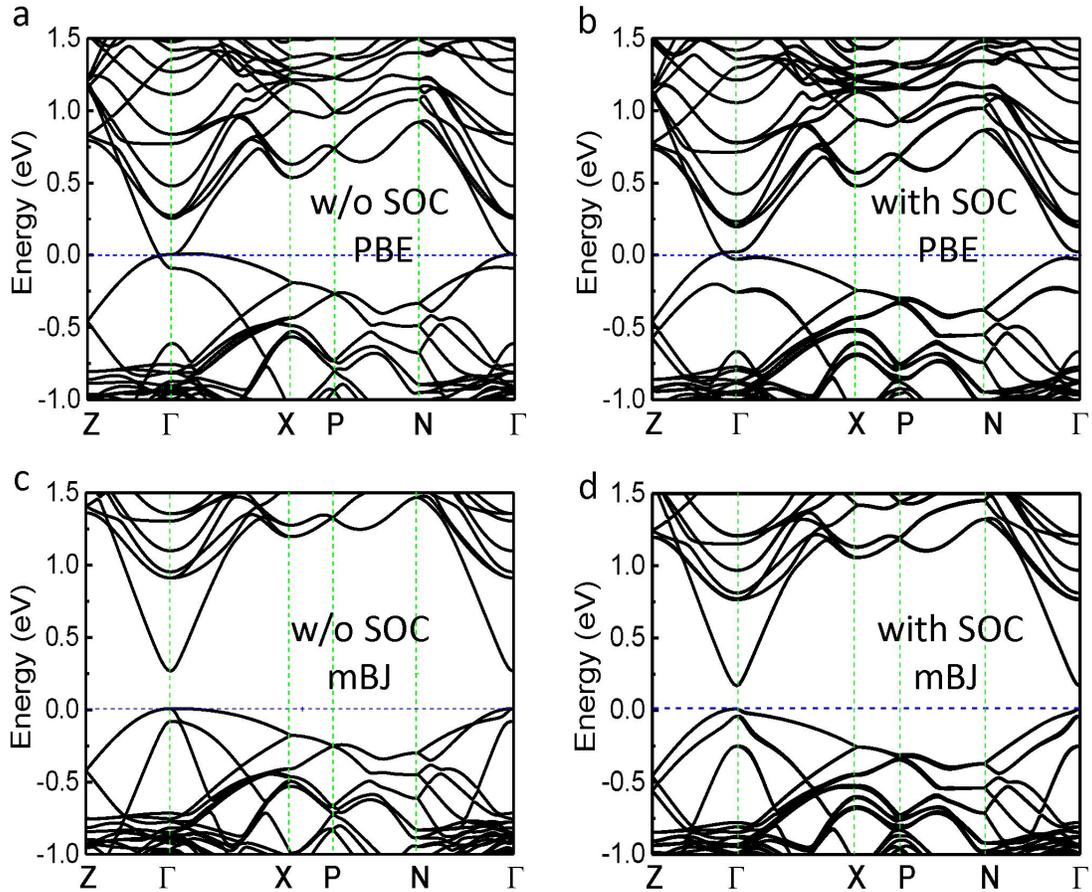

**Fig. 1.** (a) and (b) Calculated band structures for $Cd_3As_2$ by using PBE without and with SOC considered, respectively. (c) and (d) Calculated band structures for $Cd_3As_2$ by using mBJ without and with SOC considered, respectively.

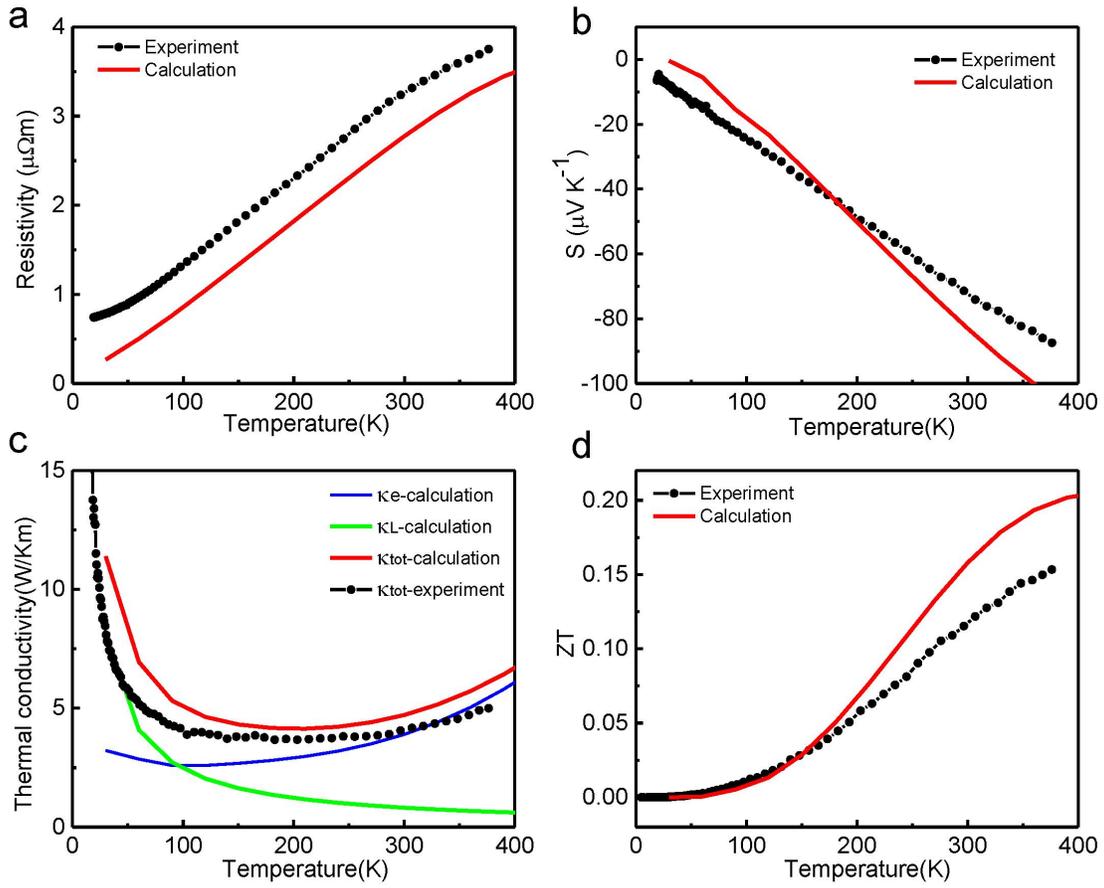

**Fig. 2.** (a)-(d) Calculated temperature-dependent resistivity, Seebeck coefficient (*S*), thermal conductivity, and figure of merit (*ZT*) of $Cd_3As_2$, compared with the experimental data from Ref. [11].

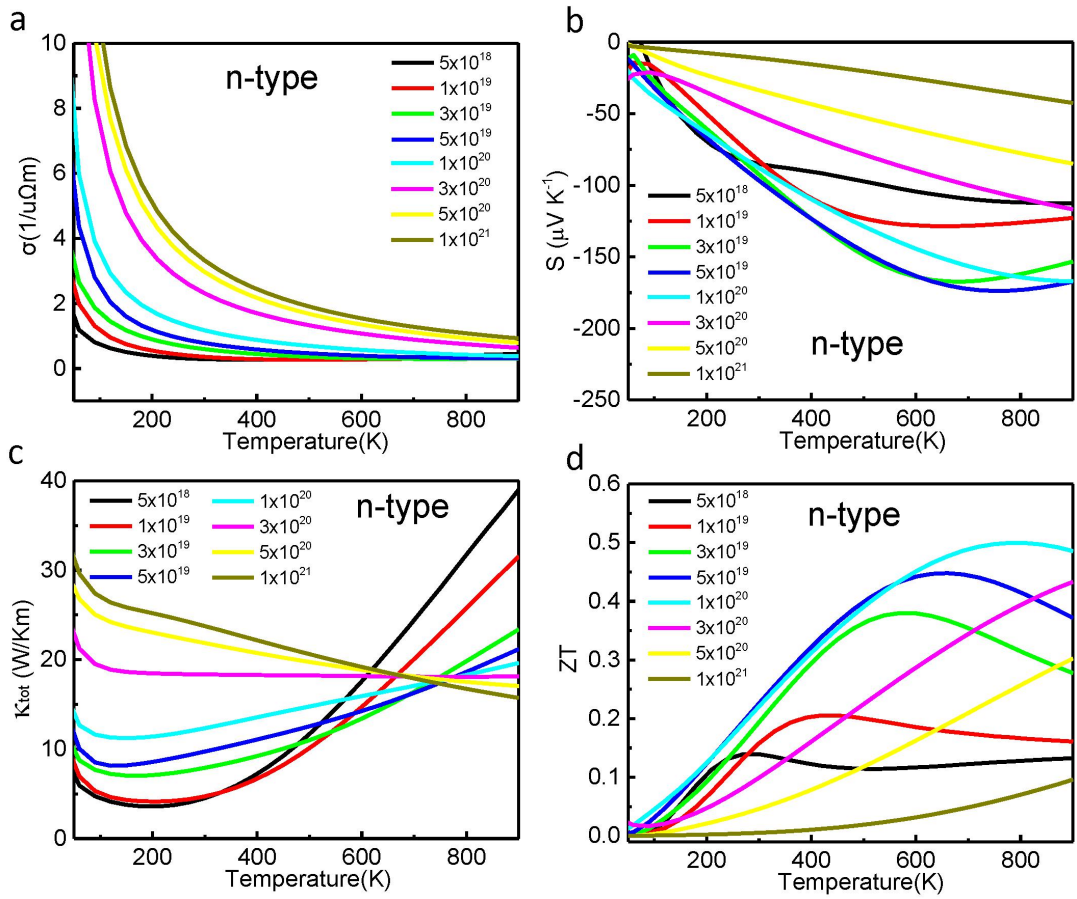

**Fig. 3.** (a)-(d) Calculated electrical conductivity (σ), Seebeck coefficient (*S*), thermal conductivity (κ), and figure of merit (*ZT*) of n-type $Cd_3As_2$ with respect to the temperature for different carrier concentrations (cm$^{-3}$).

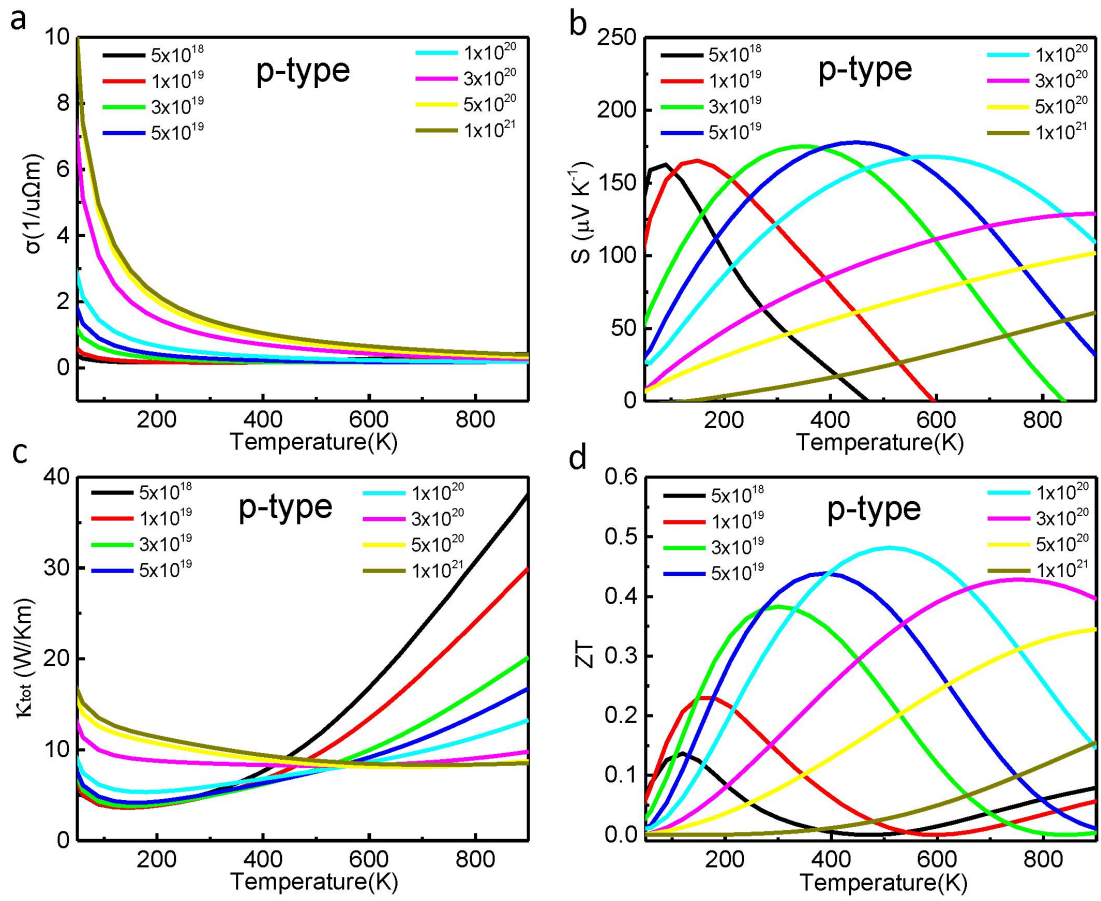

**Fig. 4.** (a)-(d) Calculated electrical conductivity (σ), Seebeck coefficient (*S*), thermal conductivity (κ), and figure of merit (*ZT*) of p-type $Cd_3As_2$ with respect to the temperature for different carrier concentrations ($cm^{-3}$).